\documentclass{cernrep}
\usepackage{amsmath,amssymb,amsfonts,color,cite}
\usepackage{placeins}

\usepackage[pdftex]{graphicx}
\pdfoutput=1 

\usepackage{eucal,mathrsfs}

\usepackage[normalem]{ulem}

\newcommand{\be}{\begin{equation}}
\newcommand{\ee}{\end{equation}}
\newcommand{\bea}{\begin{eqnarray}}
\newcommand{\eea}{\end{eqnarray}}

\newcommand{\ifb}{fb$^{-1}$}
\newcommand{\ipb}{pb$^{-1}$}

\newcommand{\tev}{\,\, \mathrm{TeV}}
\newcommand{\gev}{\,\, \mathrm{GeV}}

\title{LHC-7 supersymmetry search interpretation within the pMSSM} 

\author{Shehu~S.~AbdusSalam~\footnote{email: shehu@ictp.it}}
\institute{The Abdus Salam International Centre for Theoretical Physics,
  Strada Costiera 11, Trieste 34014, Italy}

\begin{document}

\maketitle 

\noindent

\begin{abstract}
The ATLAS collaboration published supersymmetry limits based on up to
about $4.7$~\ifb 
data collected over the year 2011 from LHC runs at $7 
\tev$. These were mainly interpreted within restricted, particular or
simplified models for supersymmetry breaking schemes or scenarios. The
pMSSM is an alternative and more generic supersymmetry framework which
captures broader phenomenological features. Searching for more generic
conclusions from the supersymmetry limits interpretation, we update a
Bayesian global fit of the pMSSM to pre-LHC data using the LHC-7
limits. The posterior distributions show the most up to date features,
revealing allowed versus excluded regions in sparticle mass
planes within the MSSM. 
\end{abstract}


\section{Introduction}
\label{sec:intro}
The discovery of Higgs boson-like state around 126.5
GeV~\cite{:2012gk} or 125 GeV~\cite{:2012gu} at the large
hadron collider (LHC) is an excellent accomplishment. However this 
only marks the beginning of exciting moments for particle physics 
endeavour in establishing the mechanism of electroweak symmetry
breaking and for shedding light on new physics beyond the Standard
Model (BSM.) During the year 2011 run of the LHC machine both ATLAS
and 
CMS detectors have recorded up to about 5~\ifb of data. The
collaborations have conducted many analyses on the data, searching 
for, amongst other new physics models, supersymmetry (SUSY) by looking
for final states containing jets and large missing transverse energy (MET)
that could indicate the production of squarks and gluinos in the
collider~\cite{ATLAS,CMS}. There is no sign for SUSY observed as of
the time of writing this article according to public results. As such 
findings of the experiments were presented in the form of  
model-independent non-SM cross section limits and interpretations showing
exclusion regions within specific models of SUSY such as the
constrained version of the R-parity conserving minimal supersymmetric
standard model (MSSM) called CMSSM/mSUGRA, particular SUSY breaking
schemes (for a review see e.g.~\cite{Nath:2003zs}) or simplified
models~\cite{Alves:2011wf} of SUSY scenarios. 

\subsection{Aim of the article}
In this article we will assess the impact of the LHC-7 SUSY results on
the R-parity conserving phenomenological MSSM (pMSSM). In particular
we are going to use the ATLAS SUSY limits reported 
in Refs.~\cite{daCosta:2011qk, Aad:2011xm, ATLAS-CONF-2011-090,
ATLAS-CONF-2011-098, Aad:2011ib, Aad:2011qa, ATLAS-CONF-2012-033,
ATLAS-CONF-2012-037, ATLAS-CONF-2012-041, ATLAS:2012ah, Aad:2012cw}
to discriminate between allowed and excluded regions in the pMSSM
sparticle mass planes. To date several research groups have looked
into analysing the impact of LHC-7 data on various SUSY models. For a
non exhaustive instances see Refs.~\cite{Balazs:2012qc,
  Allanach:2011qr, Buchmueller:2011sw,  
Bechtle:2012zk, Sekmen:2011cz, CahillRowley:2012cb, Carena:2012he,
Arbey:2011un, Cabrera:2011ds, Dolan:2011ie, Grellscheid:2011ij,
Fowlie:2011mb, Strege:2011pk, Essig:2011qg, Kats:2011qh, Brust:2011tb,
Papucci:2011wy}. Out of these, the interpretation, within the pMSSM,
for the CMS collaboration's SUSY limits based on 1~\ifb data presented
in~\cite{Sekmen:2011cz} has a particular relevance. Our analyses
updates the impact of LHC results on the pMSSM, here by using SUSY
limits from the ATLAS collaboration with up to about 4.7~\ifb data. 
Following the approach in Ref.~\cite{Sekmen:2011cz}, the data $d$ from
the various experiments used for our analysis is decomposed into two
independent parts, \be d = d_{pre-LHC} \, \oplus \, d_{LHC} \ee
where $d_{pre-LHC}$ represents the indirect pre-LHC collider and cold
dark matter relic density constraints summarised in Table~\ref{tab:obs} and
$d_{LHC}$, the LHC-7 SUSY limits shown in
Table~\ref{fig:acceptances}. The pMSSM posterior distributions  
from the Bayesian global fit~\cite{AbdusSalam:2009qd} to pre-LHC data
is now used as the prior, $\pi(\theta)$, for updating the pMSSM
posterior sample with the LHC SUSY limits. Using Bayes theorem,
weighing the prior with the likelihood over the LHC data
$L(d_{LHC}|\theta)$ gives an updated, post-LHC, posterior distribution 
\be \label{tget} p(\theta|d_{LHC}) \sim L(d_{LHC}|\theta) \, \pi(\theta) \ee 
valid up to a normalisation factor. Analysing this will reveal
information about the impact of the LHC-7 data on the pMSSM parameter
space. 

This article is structured as follows. In remaining part of this 
Section we give a brief recapitulation of the Bayesian global fit of
the pMSSM to the pre-LHC data allowing us to set the context for
explaining our analysis. In Section~\ref{sec:analysis} we
present the methodology employed in simulating SUSY event at the LHC
and the computation of the pMSSM predictions for the cross section
within acceptance, the main observable to compare with the upper
limits from ATLAS. The impact of the experimental limits on the pMSSM
is presented in Section~\ref{sec:impact}. The last Section is
preserved for discussing our conclusions and outlook. 

\subsection{The pre-LHC pMSSM fit review}
\label{sub:pmssm}
The Bayesian global fit of the pMSSM to pre-LHC data were performed
in~\cite{AbdusSalam:2008uv,AbdusSalam:2009qd}. The posterior samples
reveal SUSY spectra with various characteristics satisfying different
phenomenological scenarios~\cite{AbdusSalam:2010qp,
  AbdusSalam:2011hd,AbdusSalam:2011fc} which are mainly difficult to
capture within the classic constrained benchmark models. 
For the pMSSM fit, the parametrisation is
completely decoupled from the details of the physics responsible for
SUSY breaking.   
Requiring compatibility with observations about CP violation and
flavour changing neutral current processes, only real soft SUSY
breaking terms are considered, with all 
off-diagonal elements in the sfermion mass terms and trilinear
couplings set to zero, and the first-and second-generation soft terms
equalised; leading to a set of 20 parameters:  
\be
\theta = \{ M_{1,2,3};\;\; m^{3rd \, gen}_{\tilde{f}_{Q,U,D,L,E}},\;\; 
m^{1st/2nd \, gen}_{\tilde{f}_{Q,U,D,L,E}}; \;\;A_{t,b,\tau,\mu=e},
\;\;m^2_{H_{u,d}}, \;\;\tan \beta \}, 
\ee
where $M_1$, $M_2$ and $M_3$ are the gaugino mass parameters; and 
$m_{\tilde f}$ are the sfermion mass parameters. $A_{t,b,\tau,\mu=e}$
represent the trilinear scalar couplings while the Higgs-sector
parameters are specified by the two Higgs doublet masses $m^2_{H_1}$,
$m^2_{H_2}$, the ratio of the vacuum expectation values $\tan
\beta=\left<H_2\right>/\left<H_1\right>$ and the sign of the Higgs
doublets mixing parameter, $sign(\mu).$ 
The pre-LHC data, $d_{pre-LHC} = \underline{\delta} = \{ \mu_i,
\sigma_i \}$ where $i = 
1,2,\, \ldots$, represents the experimental central values and
errors for the electroweak physics observables, B-physics observables
and the cold dark matter relic density, summarised in
Table~\ref{tab:obs}, 
\bea
\underline O &= &\{ m_W,\; \sin^2\, \theta^{lep}_{eff},\; \Gamma_Z,\;
\delta 
a_{\mu},\; R_l^0,\; A_{fb}^{0,l},\; A^l = A^e,\; R_{b,c}^0,\;
A_{fb}^{b,c},\; A^{b,c}, \\ \nonumber 
& &BR(B \rightarrow X_s \, \gamma),\; BR(B_s \rightarrow \mu^+ \, \mu^-),\; 
\Delta_{0-},\; R_{BR(B_u \rightarrow \tau \nu)},\; R_{\Delta M_{B_s}},\\ \nonumber 
& &\Omega_{CDM}h^2 \}. 
\eea
The pre-LHC posterior distribution from the Bayesian global fit which
is now considered as a prior distribution for the update with the LHC
data is given by 
\be \pi(\theta) = \pi^\prime(\theta) \, \prod_{i=1} \, (2\pi 
\sigma_i^2)^{-1/2} \, \exp\left[-(O_i - \mu_i)^2/2 \sigma_i^2\right]
\ee  
where $\pi^\prime(\theta)$ is the prior probability density for the
Bayesian global fits to $d_{pre-LHC}$ which can be flat over the
individual parameters $\theta_i$ (for flat prior fits) or flat over
the logarithm of the parameters, $\log \theta$ (for log prior fits). Here
we do not aim at checking the constraining strength of the LHC data
over the pMSSM parameters. Thus no prior dependence analysis is
discussed and we work with only the log prior posterior-samples of the
pre-LHC global fits. In the next Section we describe the data,
i.e. the extra-SM cross sections within acceptance, the 
simulation of SUSY events at the LHC, and the ATLAS-like analyses
of the events. These are used in constructing the likelihood,
$L(d_{LHC}|\theta)$ required in addition to the prior, $\pi(\theta)$,
for completing the required variables in the target eq.(\ref{tget}).
\begin{table}
\begin{center}{\begin{tabular}{|cl||cl|}
\hline
Observable & Constraint & Observable & Constraint  \\ 
\hline
$m_W$ [GeV]& $80.399 \pm  0.027$ &$A^l = A^e$& $0.1513 \pm
0.0021$  \\
$\Gamma_Z$ [GeV]& $2.4952 \pm 0.0025$ &$A^b$ & $0.923 \pm 0.020$\\
$\sin^2\, \theta_{eff}^{lep}$  & $0.2324 \pm 0.0012$ &$A^c$ & $0.670 \pm 0.027$ \\  
$\delta a_\mu $ & $(30.2 \pm 9.0) \times 10^{10}$ &$Br(B\rightarrow
X_s \gamma)$ & $(3.55 \pm 0.42) \times 10^{4}$ \\  
$R_l^0$ & $20.767 \pm 0.025$ &$Br(B_s \rightarrow \mu^+ \mu^-)$ & $<
5.8 \times 10^{-8}$ (see caption.)\\  
$R_b^0$ & $0.21629 \pm 0.00066$ &$R_{\Delta M_{B_s}}$ & $0.85 \pm 0.11$\\ 
$R_c^0$ & $0.1721 \pm 0.0030$ &$R_{Br(B_u \rightarrow \tau \nu)}$&
$1.26 \pm 0.41$ \\ 
$A_{\textrm{FB}}^b$ & $0.0992 \pm 0.0016$ &$\Delta_{0-}$ & $0.0375 \pm
0.0289$\\  
$A_{\textrm{FB}}^c$ & $0.0707 \pm 0.035$ &$\Omega_{CDM} h^2$ & $0.11
\pm 0.02 $ \\ 
\hline
\end{tabular}}\end{center}
\caption{Summary for the central values and errors for the electroweak
  physics observables, B-physics observables and cold dark matter
  relic density constraints. The posterior distribution from the
  pre-LHC fit for $Br(B_s \rightarrow \mu^+ \mu^-)$ is centred around
  $2.8 \times 10^{-9}$ so most of the points are in agreement with the
  more recent bound $4.5 \times 10^{-9}$~\cite{Aaij:2012ac}.} 
\label{tab:obs}
\end{table}

\section{Analysis}
\label{sec:analysis}
Our analyses is centred around computing the likelihood for $d_{LHC}$,
given the pMSSM parameters, $\theta_i$. In this Section we describe the
ATLAS data and the simulation of SUSY events. The latter allows for 
computing the extra-SM, here SUSY, cross sections within
acceptances, $\sigma^{acc}$, over various cuts on the collider final
state characteristics. The degree of agreement or deviation between
the pMSSM predictions for $\sigma^{acc}$ and the experimental values
is used to quantify the plausibility of obtaining the data from the
model parameters, $L(d_{LHC}|\theta)$.

\subsection{The data, $d_{LHC}$}
The LHC data we use are the 95\% C.L. upper limits on the non-SM cross
section within acceptance. For each of the ATLAS analyses~\cite{daCosta:2011qk,
ATLAS-CONF-2011-098,Aad:2011ib, Aad:2011qa, ATLAS-CONF-2012-033,
ATLAS-CONF-2012-037, ATLAS:2012ah, Aad:2011xm, ATLAS-CONF-2011-090,
Aad:2012cw, ATLAS-CONF-2012-041}, there are various
signal regions defined by specific set of cuts and events selection
criteria. 
The results are based on various, namely 35 \ipb up to 4.7 \ifb, data set
recorded by the ATLAS detector at 7 TeV centre of mass
energy run of the LHC in the year 2011. 
The analyses were designed to capture different possible
manifestations of SUSY after the proton-proton collisions at the LHC.

SUSY production at the collider would be dominated by large direct 
production of squark and gluino pairs ($\tilde{g}\,\tilde{g}$,
  $\tilde{g}\,\tilde{q}$, or $\tilde{q}\,\tilde{q}$) 
that would decay ($\tilde{q}\rightarrow q \tilde{\chi}_1^0$ and
$\tilde{g}\rightarrow q\, \tilde{q} \, \tilde{\chi}_1^0$) to the
weakly interacting neutralino lightest supersymmetric particle (LSP),
$\tilde{\chi}_1^0$, which escapes the detector unseen in the form of 
missing transverse energy, MET. The different groups of
search channels, which we briefly describe here, are all MET-based. 
The first is the search for squarks and gluinos that lead to final states
containing high-$p_T$ jets, MET and no leptons (electrons or
muons) as in
Refs.~\cite{daCosta:2011qk,Aad:2011ib,Aad:2011qa,ATLAS-CONF-2012-033,ATLAS-CONF-2012-037}.
The strategy for this group of searches is optimised for maximal
discovery reach in the $m_{\tilde g}$-$m_{\tilde q}$ plane. 
This group of search channels could be specialised to the case of
having heavy flavour jets. Doing this would capture the scenario where
the sbottoms ($\tilde{b}_1$) and stops ($\tilde{t}_1$) are lighter
than other squarks such that direct or gluino-mediated production
($\tilde{g} \rightarrow b \tilde{b}$ or $\tilde{g} \rightarrow t
\tilde{t}$) is the dominant SUSY production mode in the collider as
considered in Refs.~\cite{ATLAS-CONF-2011-098,ATLAS:2012ah}.  

Requiring final states containing one (or more) electron or muon 
in addition to jets and MET would capture scenarios where gluinos
cascade decay products include a charginos, $\tilde{\chi}^{\pm}$,
which subsequently decays into final states containing high-$p_T$
leptons as considered in 
Refs.~\cite{ATLAS-CONF-2011-090,ATLAS-CONF-2012-041}.    
Further, in scenarios, such as Natural SUSY~\cite{Chan:1997bi}, where first and 
second generation sfermion masses are larger than few TeVs, the direct 
production of weak gauginos may be the dominant SUSY processes. When
both gauginos decay leptonically, a distinctive signature with no
jets, three leptons and significant MET, as considered in
Ref.~\cite{Aad:2012cw}, can be observed.  

In all, we considered 55 SUSY signal regions. 
For each region the number of events that pass the selected criteria
and also the expected Standard Model (background) events were
reported~\footnote{No detector simulation is attempted as that would
make no significant effect on our findings or conclusions.}. In addition
the upper bound on the cross sections for non-SM interactions were
also given. These allow for comparisons with 
any BSM predictions to determine whether the new physics model is
allowed or ruled out at the 95\% confidence level. This is the
approach we have chosen. The LHC data,
$d_{LHC}$, for our analysis is represented by the set
summarised in Table~\ref{fig:acceptances}. The limits are used to
constrain the pMSSM $\sigma^{acc}$predictions from the simulation of
SUSY production at the LHC. 

\begin{table}
  \begin{center}
    \begin{tabular}{|c|l|l|l|l|l|l|l|l|l|l|l|l|}
\hline
Channels + MET&\multicolumn{11}{|c|}{Signal regions, $\sigma_{BSM}$ upper limits}&Luminosity \\
\hline
jets + 0-lep&1.3&0.35&1.1&0.11& - & -& -& -& -& -& -&  35\,\ipb~\cite{daCosta:2011qk} \\
\hline
jets + 0-lep& 22 & 25 & 429 & 27&17& -& -& -& -& -& - & 1.04\,\ifb~\cite{Aad:2011ib} \\ 
\hline
$\geq$6 jets& 194&8.4& 12.2&4.5&-&-&-&-&-&-&-&1.34\,\ifb~\cite{Aad:2011qa} \\ 
\hline
0jet + 2-lep& 0.22& 0.09& 0.21& 0.07& 0.07& 0.07&-&-&-&-&-&35\,\ipb~\cite{Aad:2011xm} \\ 
\hline
jets + 1-lep& 41&53&-&-&-&-&-&-&-&-&-&165\,\ipb~\cite{ATLAS-CONF-2011-090} \\ 
\hline
b-jet + 0-lep&288& 61& 78& 17&-&-&-&-&-&-&-&830\,\ipb~\cite{ATLAS-CONF-2011-098}\\
\hline
$\geq$(2-6)jets + 0-lep&0.62& 5.3&6.2&0.65&3.5&3.7&12&2.2&2.6&2.5&18&4.7\,\ifb~\cite{ATLAS-CONF-2012-033} \\
\hline
$\geq$(6-9)jets + 0-lep&14& 4.2& 1.2& 9.8&
3.2&0.81&-&-&-&-&-&4.7\,\ifb~\cite{ATLAS-CONF-2012-037}  \\
\hline
$\geq$(2-4)jets + 1-lep&1.3& 1.5& 3.7&-&-&-&-&-&-&-&-&4.7\,\ifb~\cite{ATLAS-CONF-2012-041}\\ 
\hline
$\geq$(1-2)b-jets + 0-lep&283& 65& 15.4& 61& 14.4& 4.3& 22.2&
8.5&-&-&-&2.05\,\ifb~\cite{ATLAS:2012ah}\\ 
\hline
3-lep&3.5&1.5&-&-&-&-&-&-&-&-&-&2.06\,\ifb~\cite{Aad:2012cw}\\ 
\hline
    \end{tabular}
    \caption{
      The ATLAS 95\% C.L. upper limits on the extra-SM cross sections
      within acceptance for the various signal regions described in
      the text. The limits on each search channel row are
      ordered; with the first representing the first corresponding name
      of the signal region described in the corresponding experimental
      paper. The unit for each cross section is the inverse of the  
      corresponding luminosity.
      }
    \label{fig:acceptances}
  \end{center}
\end{table}

\subsection{The pMSSM predictions for the cross section within
      acceptance, $\sigma_i^{acc}$} 
In order to compute the predictions for $\sigma_i^{acc}$ within
acceptances over the cuts and selections criteria that defines the
various ATLAS SUSY signal regions, we simulate the generation of SUSY
events at 7 TeV LHC using Monte Carlo events generator and then
analyse the collider final states in a similar fashion to the ATLAS
procedures.  

\subsubsection{SUSY events simulation}
\label{simulations}
We use \texttt{Herwig++}~\cite{Bahr:2008pv,Gieseke:2011na} to simulate four sets of
sparticle production processes for each point in the pMSSM sample, namely:
a) squark-squark and squark-gluino production, 
b) the production of an electroweak gaugino in association with a
squark or gluino 
and c) the production of slepton 
and electroweak gaugino pairs. 
Each of the pMSSM posterior sample point in the SUSY Les Houches
Accord (SLHA) file
format produced by \texttt{SoftSUSY}~\cite{Allanach:2001kg} is passed
to \texttt{Herwig++}~\cite{Bahr:2008pv,Gieseke:2011na} for generating
1000 SUSY events. Through out the analysis we use the SUSY production cross
section from the event generator calculated at leading order in
perturbative QCD~\footnote{About half of the posterior
samples we consider have a negative gluino mass, allowed by the SLHA
accord, that crashes the NLO calculator we have access to. As such
the NLO correction is dropped out.}.

\subsubsection{Analysing the simulated SUSY events using Rivet}
\label{rivetanalyses}
The analysis of the generated SUSY events are done at the particle level
using the ``Robust Independent Validation of Experiment and Theory'',
\texttt{Rivet}, Monte Carlo validation
framework~\cite{Buckley:2010ar}. We use this to analyse each and every
SUSY event generated by the Monte Carlo collider simulator, without
the need for detector simulations, and the publicly available Rivet
analyses for the ATLAS SUSY searches in 
Refs.~\cite{daCosta:2011qk, 
  Aad:2011xm, ATLAS-CONF-2011-090, ATLAS-CONF-2011-098, Aad:2011ib,
  Aad:2011qa, ATLAS-CONF-2012-033, ATLAS-CONF-2012-037,
  ATLAS-CONF-2012-041, ATLAS:2012ah, Aad:2012cw}. The Rivet analyses
are plugged-in to \texttt{Herwig++} for computing the acceptances,
$A_i = N_{cuts}/N_{total}$, after applying the various cuts on the
kinematic variables of the collider final states. Here $N_{cuts}$ is
the number of events that pass the experimental cuts and
$N_{total}=1000$ is the total number of generated SUSY events. Rivet
acts per-event wise on 
the events produced by \texttt{Herwig++}. The jet identification is done
using \texttt{fastjet}~\cite{Cacciari:2011ma}. 
The cross section within acceptance is computed as 
\be
\sigma^{acc}_{i} =  \epsilon \, A_i \, \sigma^{SUSY, LO}_{i} 
\ee
where we consider the efficiency, $\epsilon = 1$ (since no detector
simulation\footnote{Performing the analysis with fast detector
  simulation will not have a significant effect on our results or
  conclusions. For instance, a $\pm 2 \gev$ accuracy gain in the MET
  or jet $p_T$ will have little or no affect on our results.});
$i=1,2,\dots,55$ runs over the 55 different signal regions, and
$\sigma^{SUSY, LO}$ is the total LO SUSY production cross section. 

\subsection{The likelihood, $L(d_{LHC}| \theta)$} 
The likelihood  is a simple step function that equals to unity if
the ATLAS limits are satisfied else zero if excluded. SUSY points
with predicted cross sections smaller (greater) than the ATLAS non-SM
limits are then allowed (excluded) at 95\% confidence level according
to 
\be
L(d_{LHC}|\theta) = \prod_{i=1}^{55} \ell_{i}; \quad \ell_{i} = 
\left\{ 
\begin{array}{ll}
  0 & \textrm{if } \sigma^{acc}_{i} > \sigma^{acc,max}_{i}\\
  1 & \textrm{if } \sigma^{acc}_{i} \leq \sigma^{acc,max}_{i}
\end{array} \right\}.
\ee

\section{The SUSY limits' impact on the pMSSM}
\label{sec:impact}
\begin{table}
  \begin{center}
    \begin{tabular}{|c|l|l|}
      \hline
      No. &     Constraint &            Log prior survive  \\
      \hline
      1.&      $m_h=122-128 \gev$&         36.08\%\\
      2.&      $m_h=125-126.5 \gev$&       9.17\%\\     
      3.& SUSY $\sigma^{acc}$ limits & 57.47\% \\
      4.& $m_h(1)$ and SUSY $\sigma^{acc}$ limits & 14.53\%\\ 
      5.& 1.67-$\sigma$ $R_{ZZ ; \gamma\gamma}$ & 13.30\%\\           
      6.& $m_h(1)$ \& 1.67-$\sigma$ $R_{ZZ ; \gamma\gamma}$ & 4.67\%\\           
      7.& 1.67-$\sigma$ $R_{ZZ ; \gamma\gamma}$ \& SUSY
      $\sigma^{acc}$ limits  & 6.11\%\\ 
      8.& $m_h$, 1.67-$\sigma$ $R_{ZZ ; \gamma\gamma}$ \& SUSY
      $\sigma^{acc}$ limits  & 2.19\%\\ 
      \hline 
    \end{tabular}
    \caption{Summary of the relative number of surviving posterior points,
      from the pre-LHC Bayesian global fits of the pMSSM, and after imposing
      the Higgs discovery data and SUSY limits.} 
    \label{tab:summary}
  \end{center}
\end{table}

The relative number of the pMSSM points that survive the SUSY limits
and in various combinations with the Higgs-sector data (in the
di-photon channels as applied in Ref.~\cite{AbdusSalam:2012sy} and
other ATLAS, CMS, LEP and Tevatron Higgs-sector constraints
implemented in the HEP packages
\texttt{HiggsBounds}~\cite{Bechtle:2008jh} and
\texttt{FeynHiggs}~\cite{Degrassi:2002fi}) are summarised in  
Table~\ref{tab:summary}. The posterior probability 
distribution for the sparticle masses derived from the post-LHC
distribution eq.(\ref{tget}) are shown in Fig.\ref{fig:masses}. 
As can be seen from the mentioned plots, the ATLAS SUSY limits are
most sensitive in constraining the gluino mass whose central value is
now shifted\footnote{The SUSY search channels that involves final states
  with high number of jet multiplicities severely constrains the gluino
  mass to higher values relative to the pre-LHC mass
  distribution.} from around $2 \tev$ to $3 \tev$ followed by the squark
masses. The effect of the SUSY limits on the gluino mass seems to be
very different from what happen for the case of constrained models
such as CMSSM/mSUGRA (addressing this is outside the purpose of this
article). The limits are relatively less sensitive in constraining 
sleptons, electroweak gauginos and the lighter sbottom or stop quarks.  
The resultant effect of applying the limits coming from the 55
different ATLAS' SUSY signal regions shown in
Table~\ref{fig:acceptances} on the pMSSM can be summarised in plots
showing allowed versus ruled out regions in sparticle mass planes. For
gluino-stop and gluino-sbottom mass planes the plots are shown
in Fig.~\ref{ExclC}. Region where
$max(\frac{\sigma_i^{acc}}{\sigma_i^{acc,max}}) > 1.0$ are excluded at
95\% confidence level by the combined ATLAS limits. The plots shows
that unlike the gluino mass which is now constrained to be heavy $\sim
3 \, TeV$, the third generation squarks can be much lighter.

\begin{figure*}
  \begin{center}
    \vspace{-6cm}
    \hspace{-4.cm}
    \includegraphics[width=1.25\textwidth]{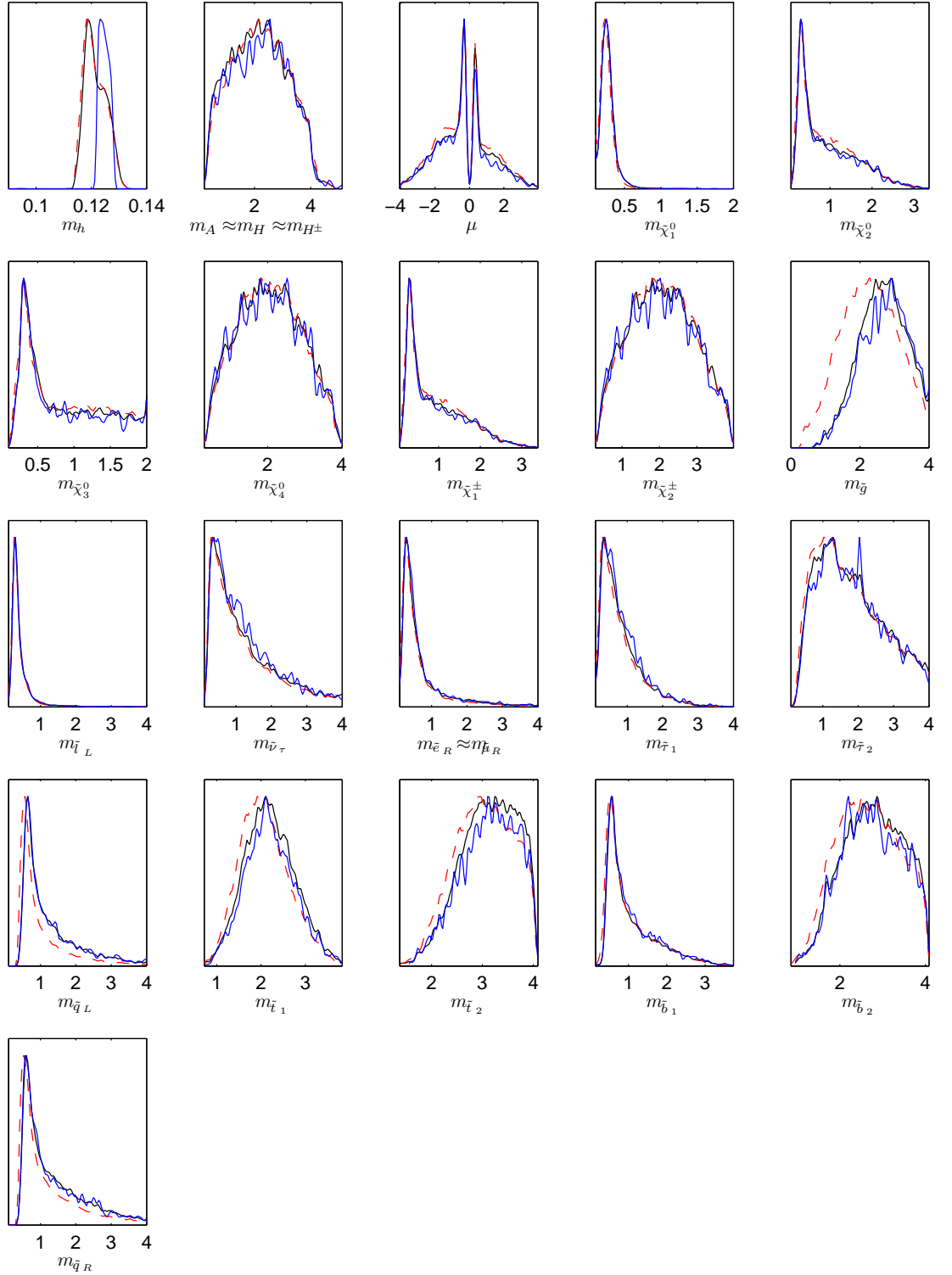} 
  \end{center}
  \vspace{-2.4cm}
  \caption{The plots compare the log prior pMSSM sparticle
    masses' marginalised 1-dimensional pre-LHC posterior distributions 
    (dashed-red curves) and the surviving parameter regions after
    imposing only the SUSY limit (black curves) and both $m_h = 122.0 - 128.0 
    \gev$ and SUSY limits together (blue curves). All the masses are 
    in $\tev$ units. The vertical axes represent the relative
    probability weights of the model points.} 
  \label{fig:masses}
\end{figure*}

\begin{figure}
  \begin{minipage}[b]{10.5cm}
    \centering
    \hspace{-3cm} 
    \includegraphics[width=10cm]{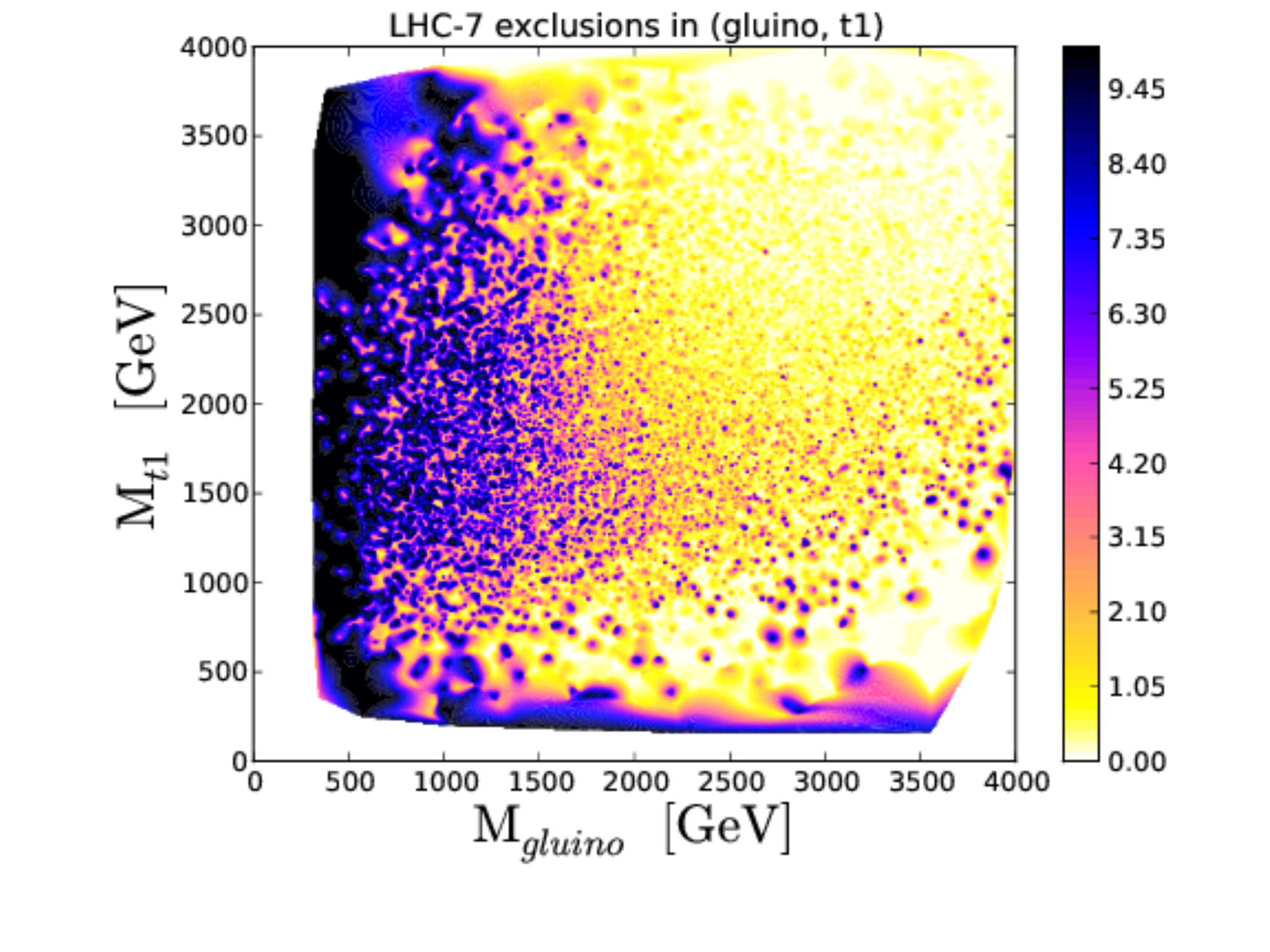}
  \end{minipage}
  \hspace{-3.8cm} 
  \begin{minipage}[b]{10.5cm}
    \centering
    \includegraphics[width=10cm]{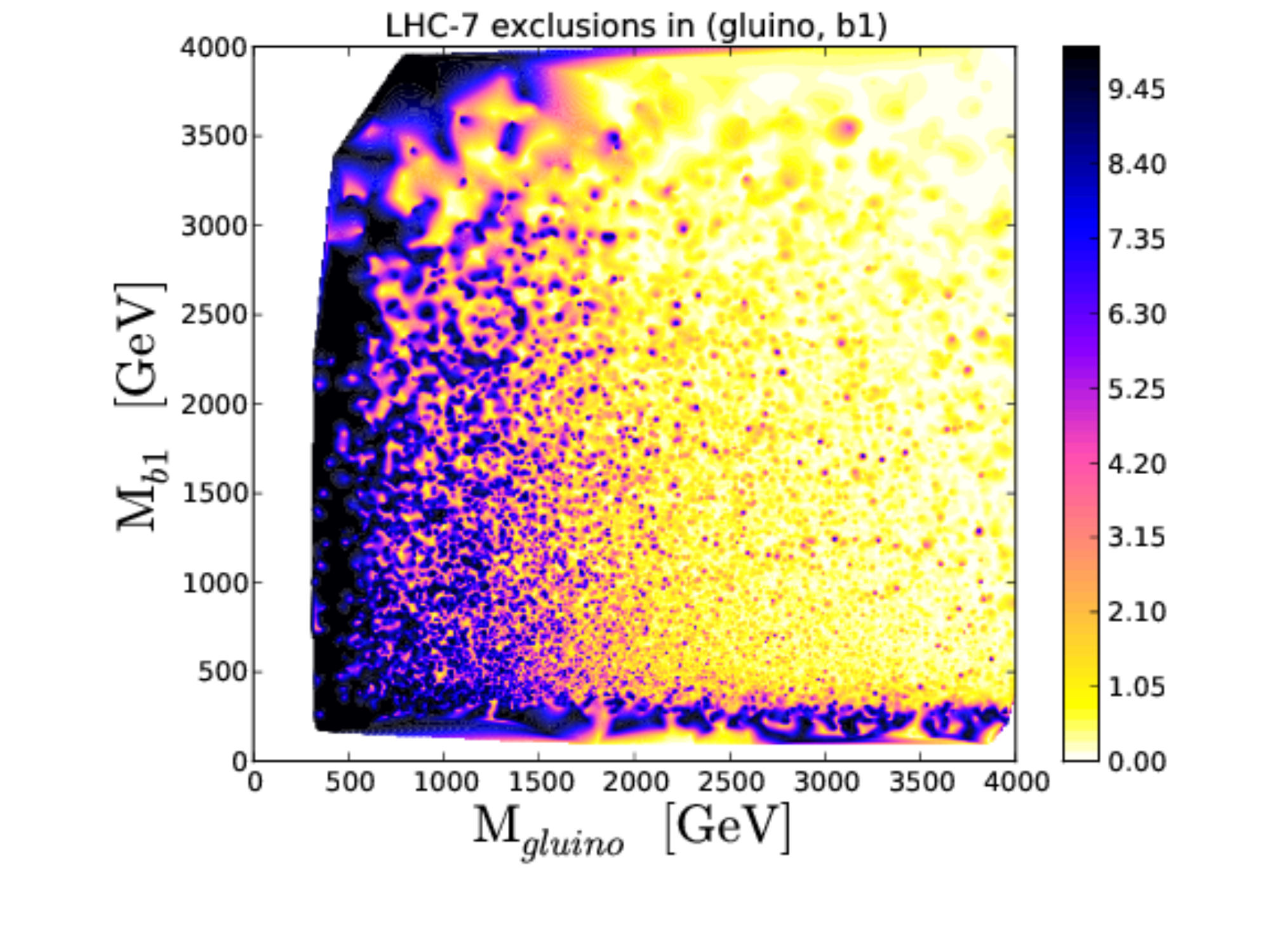}
  \end{minipage}
  \vspace{-1.5cm}
  \caption{The plots show 95\% confidence level exclusion contours for 
    gluino-stop and gluino-sbottom  mass planes derived from the combined set
    of ATLAS limits. The colour scales are
    proportional to the expected number of signal events normalised to
    the combined exclusion limit. The contour
    $max(\frac{\sigma^{acc}}{\sigma^{cutoff}}) = 1.0$ determines the
    exclusion boundaries within the sparticle masses plane. Region with
    colour code greater than unity are excluded at 95\% confidence
    level.} 
  \label{ExclC} 
\end{figure}

\section{Conclusions and outlook}
We have computed the effect of SUSY searches results from ATLAS
collaboration with up to 4.7 \ifb of 7 TeV LHC data sets taken during
the year 2011. This is done by simulating SUSY events at the LHC with
\texttt{Herwig++}~\cite{Bahr:2008pv} and the experimental analyses of
the data with \texttt{Rivet}~\cite{Buckley:2010ar}. The 
particle-level per-event Rivet analysis is used for calculating the pMSSM
predictions for the cross section within acceptances of the analyses
cuts and events selection criteria. Comparing with the
experimental 95\% C.L. upper limits (shown in
Table~\ref{fig:acceptances}) rule out about 40\% of the initial 
pMSSM sample from a pre-LHC global fit to data. The SUSY limits are
most sensitive in constraining the gluino mass whose distribution is
now centred around 3 TeV (shifted from 2 TeV.) There is also a shift,
but less significant compared to gluino mass case, in preference for
heavier 1st/2nd generation and stop masses as can be seen in the
posterior distributions in Fig.~\ref{fig:masses}. The excluded versus
allowed regions by the ATLAS' extra-SM cross section upper limits on
the gluino-stop and gluino-sbottom mass planes are shown in
Fig.~\ref{ExclC}.

Combining the SUSY and Higgs boson discovery data as described in
Ref.~\cite{AbdusSalam:2012sy} further constrain the pre-LHC posterior
samples as summarised in Table~\ref{tab:summary} with only a single
point (out of the initially about 40000 pre-LHC posterior sample
points) surviving all of the following requirements: 
\bea
\nonumber 
\textrm{SUSY } &\sigma^{acc}& \textrm{upper limits,}\\
\nonumber 
m_h &=& 125-126.5 \gev, \textrm{ and} \\ 
\nonumber 
R_{ZZ \, ; \gamma\gamma} & \equiv & \frac{\mu_{ZZ}}{\mu_{\gamma\gamma}}
= 0.56 \pm 0.25.\nonumber 
\eea  
Here $\mu_X = \frac{ \sigma(gg \rightarrow h) \, Br(h \rightarrow X)}{
  \sigma(gg \rightarrow h)_{SM} \, Br(h \rightarrow X)_{SM}}$ where $X
= \gamma \gamma \textrm{ or } ZZ$. The spectrum
 is characterised with a quasi-degenerate sbottom and
LSP and heavy gluino and 1st/2nd generation squarks. It is of the
difficult-to-see at the LHC type discussed in
Refs.~\cite{LeCompte:2011cn,AbdusSalam:2011hd}. It is worth mentioning
that the data from the Higgs sector (4.67\% model points survived) is
far more constraining compared to the SUSY limits (57.47\% model
points survived). There is however some complementarity between the
two set of constrained since applying the SUSY limits on the
Higgs-data surviving models points brings down the surviving number to
2.19\%. 

Our results go beyond the 1 \ifb analysis done in
Ref.~\cite{Sekmen:2011cz} from various perspectives. First, our
pre-LHC prior construction takes into account the constraint on the
neutralino LSP relic density. We use the more stringent ATLAS
SUSY limits from up to 4.7 \ifb of data which include channels that
better constrain gluino production with subsequent decay chains with
several jets in the final states.

The results and conclusion obtained from our analyses are 
conservative. This is because: the SUSY signal simulations were done
at leading order (LO) in perturbative QCD as opposed to 
the next-to-leading order (NLO) results reported by ATLAS
\footnote{About half of the volume of the pMSSM sample we employ have
  a negative gluino mass. This feature is the standard format allowed by
  the SLHA conventions. However such negative gluino SUSY points are not
  possible for the NLO cross section calculator that we have access
  to.}
Since the NLO cross sections are generally greater compared to the LO
values, the exclusions here are more conservative, we will be allowing 
points that otherwise will be ruled out as is the case for the
SPS4 bench mark point in Tab.~\ref{spspts} where the the LO and NLO
cross sections within acceptance for snow mass points and slopes
benchmark points can be compared.  
There is an approximate agreement, within the expected $\sim 50 \%$
accuracies, with the corresponding values obtained by previous
computations (with NLO corrections)~\cite{Dolan:2011ie}. 
\begin{table}
  \footnotesize{
  \begin{center}
    \begin{tabular}{|c|l|l|l|l|c|}
      \hline
      Benchmark point & \multicolumn{4}{|c|}{$\sigma/{\rm pb}$} & status\\\hline
      &A&B&C&D& ATLAS 35pb$^{-1}$\\\hline\hline
      ATLAS limits  & 1.3 & 0.35 & 1.1 & 0.11 & \\
      \hline\hline
      sps1a & 1.347  & 0.640  & 1.172  & 0.299  &  A,B,C,D\\
      & 2.031 & 0.933 & 1.731 & 0.418 & A,B,C,D\\
      \hline
      sps1b &0.077  &0.057  &0.062  &0.041 & allowed \\
      & 0.120 & 0.089 & 0.098 & 0.067&allowed\\
      \hline
      sps2 &0.499  &0.280  &0.425  &0.169 & D \\
      & 0.674 & 0.388 & 0.584 & 0.243 & B,D\\
      \hline
      sps3 &0.079  &0.059  &0.061  &0.043 & allowed\\
      & 0.123 & 0.093& 0.097 & 0.067 &allowed\\
      \hline
      sps4 &0.218  &0.132  &0.195  &0.084 & allowed \\
      & 0.334& 0.199 & 0.309 & 0.144 & D\\
      \hline
      sps5 &0.468  &0.259  &0.417  &0.125 & D \\
      & 0.606 & 0.328 & 0.541 & 0.190 & D\\
      \hline
      sps6 &0.523  &0.289  &0.411  &0.149 & D \\
      & 0.721 & 0.416 & 0.584 & 0.226 & B,D\\
      \hline
      sps7 &0.007  &0.005  &0.008  &0.005 & allowed \\
      & 0.022 & 0.016& 0.023 & 0.015 &allowed\\
      \hline
      sps8 &0.011  &0.005  &0.015  &0.003 & allowed\\
      & 0.021 & 0.011 & 0.022 & 0.009 &allowed\\
      \hline
      sps9 &0.015  &0.003  &0.004  &0.001 & allowed\\
      & $0.019$ & $0.004$ & $0.006$ & $0.002$ & allowed\\
      \hline
    \end{tabular}
    \caption{
      The status of SUSY snow mass points and slopes
      benchmarks~\cite{Allanach:2002nj} predicted with our LO
      calculations (top values) compared to the NLO done
      in~\cite{Dolan:2011ie} (bottom values). 
      The conservative nature of the LO results manifests for the SPS4 
      case which is rather ruled out by the NLO calculations. Note the
      agreements for the SPS9 cross sections which are all computed at
      LO.}  
    \label{spspts}
  \end{center}
  }
\end{table}

For the analysis here only the posterior samples from a pre-LHC global
fit to data with a logarithmic prior distribution over the 20 pMSSM
parameters were considered. No analysis is done with the flat prior
sample because it know that the pre-LHC fits were
prior-dependent. However, it will be interesting~\cite{wuc} to
estimate the strength of the LHC data by checking whether it allows
for prior independent results which is necessarily needed for making
conclusions regarding the predictive power of the pMSSM. This seems
possible given the apparent interplay between Higgs boson decay rate
in the di-photon decay channels which would require light
sparticles \footnote{Assuming that R-parity conserving MSSM is
  responsible for the physics behind the observed deviation from the
  SM model expectation in the di-photon channel.} and the absence of
SUSY signal to date at the LHC.

\section*{Acknowledgments}
Thanks to D.~Choudhury, F.~Quevedo, and D.~Grellscheid for helpful
comments and discussions; and to the IPPP for hospitality during the
early stage of this project.


\begin{thebibliography}{99}
\bibitem{:2012gk}
  G.~Aad {\it et al.}  [ATLAS Collaboration],
  ``Observation of a new particle in the search for the Standard Model Higgs boson with the ATLAS detector at the LHC,''
  Phys.\ Lett.\ B {\bf 716} (2012) 1.

\bibitem{:2012gu}
  S.~Chatrchyan {\it et al.}  [CMS Collaboration],
  ``Observation of a new boson at a mass of 125 GeV with the CMS experiment at the LHC,''
  Phys.\ Lett.\ B {\bf 716} (2012) 30.

\bibitem{ATLAS}
ATLAS Supersymmetry Searches,\\
{\bf https://twiki.cern.ch/twiki/bin/view/AtlasPublic/SupersymmetryPublicResults}

\bibitem{CMS}
CMS Supersymmetry Searches, \\
{\bf https://twiki.cern.ch/twiki/bin/view/CMSPublic/PhysicsResultsSUS}

\bibitem{Nath:2003zs}
  P.~Nath,
  ``Twenty years of SUGRA,''
  hep-ph/0307123.

\bibitem{Alves:2011wf}
  D.~Alves {\it et al.}  [LHC New Physics Working Group Collaboration],
  J.\ Phys.\ G {\bf 39} (2012) 105005
  [arXiv:1105.2838 [hep-ph]].

\bibitem{daCosta:2011qk}
  G.~Aad {\it et al.}  [Atlas Collaboration],
  Phys.\ Lett.\ B {\bf 701} (2011) 186.


\bibitem{Aad:2011xm}
  G.~Aad {\it et al.}  [ATLAS Collaboration],
  Eur.\ Phys.\ J.\ C {\bf 71} (2011) 1682.


\bibitem{ATLAS-CONF-2011-090}
  The ATLAS Collaboration,
  and one lepton at sqrt(s) = 7 TeV,'' 
  ATLAS-CONF-2011-090, Jun 2011.

\bibitem{ATLAS-CONF-2011-098}
  The ATLAS Collaboration,
  ATLAS-CONF-2011-098, Jul 2011.


\bibitem{Aad:2011ib}
  G.~Aad {\it et al.}  [ATLAS Collaboration],
  Phys.\ Lett.\ B {\bf 710} (2012) 67.


\bibitem{Aad:2011qa}
  G.~Aad {\it et al.}  [Atlas Collaboration],
  JHEP {\bf 1111} (2011) 099.


\bibitem{ATLAS-CONF-2012-033}
  The ATLAS Collaboration, ATLAS-2012-CONF-2012-033, Mar 2012.

\bibitem{ATLAS-CONF-2012-037}
  The ATLAS Collaboration, ATLAS-2012-CONF-2012-037, Mar 2012.

\bibitem{ATLAS-CONF-2012-041}
  The ATLAS Collaboration, ATLAS-2012-CONF-2012-041, Mar 2012.

\bibitem{ATLAS:2012ah}
  G.~Aad {\it et al.}  [ATLAS Collaboration],
  [arXiv:1203.6193 [hep-ex]].

\bibitem{Aad:2012cw}
  G.~Aad {\it et al.}  [ATLAS Collaboration],
  arXiv:1204.5638 [hep-ex].


\bibitem{Balazs:2012qc}
  C.~Balazs, A.~Buckley, D.~Carter, B.~Farmer and M.~White,
  ``Should we still believe in constrained supersymmetry?,''
  arXiv:1205.1568 [hep-ph].

\bibitem{Allanach:2011qr}
  B.~C.~Allanach, T.~J.~Khoo and K.~Sakurai,
  ``Interpreting a 1 $fb^-1$ ATLAS Search in the Minimal Anomaly Mediated
  Supersymmetry Breaking Model,''
  arXiv:1110.1119 [hep-ph].

\bibitem{Buchmueller:2011sw}
  O.~Buchmueller {\it et al.},
  ``Supersymmetry in Light of 1/fb of LHC Data,''
  arXiv:1110.3568 [hep-ph].


\bibitem{Bechtle:2012zk}
  P.~Bechtle, T.~Bringmann, K.~Desch, H.~Dreiner, M.~Hamer, C.~Hensel, M.~Kramer and N.~Nguyen {\it et al.},
  ``Constrained Supersymmetry after two years of LHC data: a global view with Fittino,''
  JHEP {\bf 1206} (2012) 098.

\bibitem{Sekmen:2011cz}
  S.~Sekmen, S.~Kraml, J.~Lykken, F.~Moortgat, S.~Padhi, L.~Pape, M.~Pierini and H.~B.~Prosper {\it et al.},
  ``Interpreting LHC SUSY searches in the phenomenological MSSM,''
  JHEP {\bf 1202} (2012) 075.

\bibitem{CahillRowley:2012cb}
  M.~W.~Cahill-Rowley, J.~L.~Hewett, S.~Hoeche, A.~Ismail and T.~G.~Rizzo,
  ``The New Look pMSSM with Neutralino and Gravitino LSPs,''
  arXiv:1206.4321 [hep-ph].

\bibitem{Carena:2012he}
  M.~Carena, J.~Lykken, S.~Sekmen, N.~R.~Shah and C.~E.~M.~Wagner,
  ``The pMSSM Interpretation of LHC Results Using Rernormalization Group Invariants,''
  arXiv:1205.5903 [hep-ph].


\bibitem{Arbey:2011un}
  A.~Arbey, M.~Battaglia and F.~Mahmoudi,
  ``Implications of LHC Searches on SUSY Particle Spectra: The pMSSM Parameter Space with Neutralino Dark Matter,''
  Eur.\ Phys.\ J.\ C {\bf 72} (2012) 1847.

\bibitem{Cabrera:2011ds}
  M.~E.~Cabrera, J.~A.~Casas, V.~A.~Mitsou, R.~Ruiz de Austri and J.~Terron,
  ``Histogram comparison as a powerful tool for the search of new physics at LHC. Application to CMSSM,''
  JHEP {\bf 1204} (2012) 133.


\bibitem{Dolan:2011ie}
  M.~J.~Dolan, D.~Grellscheid, J.~Jaeckel, V.~V.~Khoze and P.~Richardson,
  ``New Constraints on Gauge Mediation and Beyond from LHC SUSY Searches at 7 TeV,''
  JHEP {\bf 1106} (2011) 095.


\bibitem{Grellscheid:2011ij}
  D.~Grellscheid, J.~Jaeckel, V.~V.~Khoze, P.~Richardson and C.~Wymant,
  ``Direct SUSY Searches at the LHC in the light of LEP Higgs Bounds,''
  JHEP {\bf 1203} (2012) 078.

\bibitem{Fowlie:2011mb}
  A.~Fowlie, A.~Kalinowski, M.~Kazana, L.~Roszkowski and Y.~L.~S.~Tsai,
  ``Bayesian Implications of Current LHC and XENON100 Search Limits for the Constrained MSSM,''
  Phys.\ Rev.\ D {\bf 85} (2012) 075012.

\bibitem{Strege:2011pk}
  C.~Strege, G.~Bertone, D.~G.~Cerdeno, M.~Fornasa, R.~R.~de Austri and R.~Trotta,
  ``Updated global fits of the cMSSM including the latest LHC SUSY and Higgs searches and XENON100 data,''
  JCAP {\bf 1203} (2012) 030.

\bibitem{Essig:2011qg}
  R.~Essig, E.~Izaguirre, J.~Kaplan and J.~G.~Wacker,
  ``Heavy Flavor Simplified Models at the LHC,''
  arXiv:1110.6443 [hep-ph].

\bibitem{Kats:2011qh}
  Y.~Kats, P.~Meade, M.~Reece and D.~Shih,
  ``The Status of GMSB After 1/fb at the LHC,''
  arXiv:1110.6444 [hep-ph].

\bibitem{Brust:2011tb}
  C.~Brust, A.~Katz, S.~Lawrence and R.~Sundrum,
  ``SUSY, the Third Generation and the LHC,''
  arXiv:1110.6670 [hep-ph].


\bibitem{Papucci:2011wy}
  M.~Papucci, J.~T.~Ruderman and A.~Weiler,
  ``Natural SUSY Endures,''
  arXiv:1110.6926 [hep-ph].

\bibitem{AbdusSalam:2008uv}
  S.~S.~AbdusSalam,
  ``The Full 24-Parameter MSSM Exploration,''
  AIP Conf.\ Proc.\  {\bf 1078 } (2009)  297-299.

\bibitem{AbdusSalam:2009qd}
  S.~S.~AbdusSalam, B.~C.~Allanach, F.~Quevedo, F.~Feroz, M.~Hobson,
  ``Fitting the Phenomenological MSSM,''
  Phys.\ Rev.\  {\bf D81 } (2010)  095012.

\bibitem{AbdusSalam:2010qp}
  S.~S.~AbdusSalam and F.~Quevedo,
  ``Cold Dark Matter Hypotheses in the MSSM,''
  Phys.\ Lett.\ B {\bf 700} (2011) 343.

\bibitem{AbdusSalam:2011hd}
  S.~S.~AbdusSalam,
  ``Can the LHC rule out the MSSM?,''
  Phys.\ Lett.\ B {\bf 705} (2011) 331.

\bibitem{AbdusSalam:2011fc}
  S.~S.~AbdusSalam, B.~C.~Allanach, H.~K.~Dreiner, J.~Ellis, {\it et al.},
  ``Benchmark Models, Planes, Lines and Points for Future SUSY Searches at the LHC,''
  Eur.\ Phys.\ J.\ C {\bf 71} (2011) 1835.


\bibitem{Aaij:2012ac}
  R.~Aaij {\it et al.}  [LHCb Collaboration],
  Phys.\ Rev.\ Lett.\  {\bf 108} (2012) 231801
  [arXiv:1203.4493 [hep-ex]].


\bibitem{Chan:1997bi}
  K.~L.~Chan, U.~Chattopadhyay and P.~Nath,
  Phys.\ Rev.\ D {\bf 58} (1998) 096004
  [hep-ph/9710473].


\bibitem{Bahr:2008pv}
  M.~Bahr, S.~Gieseke, M.~A.~Gigg, D.~Grellscheid, K.~Hamilton,
  O.~Latunde-Dada, S.~Platzer and P.~Richardson {\it et al.},
 ``Herwig++ Physics and Manual,''
  Eur.\ Phys.\ J.\ C {\bf 58} (2008) 639.

\bibitem{Gieseke:2011na}
S.~Gieseke et.~al., 
''Herwig++ 2.5 Release Note,'' 
arXiv:1102.1672.

\bibitem{Allanach:2001kg}
B.~C. Allanach, 
''SOFTSUSY: a program for calculating supersymmetric,''
Comput. Phys. Commun. {\bf 143} (2002) 305--331.

\bibitem{Buckley:2010ar}
  A.~Buckley, J.~Butterworth, L.~Lonnblad, H.~Hoeth, J.~Monk,
  H.~Schulz, J.~E.~von Seggern and F.~Siegert {\it et al.}, 
  ``Rivet user manual,''
  arXiv:1003.0694 [hep-ph].



\bibitem{Cacciari:2011ma}
  M.~Cacciari, G.~P.~Salam and G.~Soyez,
  Eur.\ Phys.\ J.\ C {\bf 72} (2012) 1896
  [arXiv:1111.6097 [hep-ph]].


\bibitem{AbdusSalam:2012sy}
  S.~S.~AbdusSalam and D.~Choudhury,
  arXiv:1210.3331 [hep-ph].



\bibitem{Bechtle:2008jh}
  P.~Bechtle, O.~Brein, S.~Heinemeyer, G.~Weiglein and K.~E.~Williams,
  ``HiggsBounds: Confronting Arbitrary Higgs Sectors with Exclusion
  Bounds from LEP and the Tevatron,''
  Comput.\ Phys.\ Commun.\  {\bf 181} (2010) 138.

\bibitem{Degrassi:2002fi}
  G.~Degrassi, S.~Heinemeyer, W.~Hollik, P.~Slavich and G.~Weiglein,
  ``Towards high precision predictions for the MSSM Higgs sector,''
  Eur.\ Phys.\ J.\ C {\bf 28} (2003) 133.

\bibitem{LeCompte:2011cn}
  T.~J.~LeCompte and S.~P.~Martin,
  ``Large Hadron Collider reach for supersymmetric models with compressed mass spectra,''
  Phys.\ Rev.\ D {\bf 84} (2011) 015004
  [arXiv:1105.4304 [hep-ph]].

\bibitem{Allanach:2002nj}
  B.~C.~Allanach, M.~Battaglia, G.~A.~Blair, M.~S.~Carena, A.~De Roeck, A.~Dedes, A.~Djouadi and D.~Gerdes {\it et al.},
  ``The Snowmass points and slopes: Benchmarks for SUSY searches,''
  Eur.\ Phys.\ J.\ C {\bf 25} (2002) 113.

\bibitem{wuc}
  S.~S.~AbdusSalam et.~al., project under consideration. 




\end{thebibliography}
\end{document}